\documentclass[aps,pre,twocolumn,showpacs,superscriptaddress,amsmath,amssymb,floatfix]{revtex4}
\usepackage{graphicx}
\usepackage{dcolumn}
\usepackage{bm}
\usepackage{color}
\begin{document}

\title{Capillary Contact Angle in a Completely Wet Groove}

\author{A.O.\ Parry}
\affiliation{Department of Mathematics, Imperial College London, London SW7 2BZ, UK}

\author{A.\ Malijevsk\'y}
\affiliation{Department of Physical Chemistry, Institute of Chemical Technology Prague,
16628 Praha 6 Czech Republic; ICPF, Academy of Sciences, 16502 Prague 6, Czech Republic}

\author{C.\ Rasc\'{o}n}
\affiliation{GISC, Departamento de Matem\'aticas, Universidad Carlos III de Madrid, 28911 Legan\'es, Madrid, Spain}

\begin{abstract}
We consider the phase equilibria of a fluid confined in a deep capillary groove of width $L$ with identical side walls and a bottom made of a different material. All walls are completely wet by the liquid. Using density functional theory and interfacial models, we show that the meniscus separating liquid and gas phases at two phase capillary-coexistence meets the bottom capped end of the groove at a capillary contact angle $\theta^\textup{cap}(L)$ which depends on the difference between the Hamaker constants. If the bottom wall has a weaker wall-fluid attraction than the side walls, then $\theta^\textup{cap}>0$ even though all the isolated walls are themselves completely wet. This alters the capillary condensation transition which is now first-order; this would be continuous in a capped capillary made wholly of either type of material. We show that the capillary contact angle $\theta^\textup{cap}(L)$ vanishes in two limits, corresponding to different capillary wetting transitions. These occur as the width i) 
becomes macroscopically large, and ii) is reduced to a microscopic value determined by the difference in Hamaker constants. This second wetting transition is characterised by large scale fluctuations and essential critical singularities arising from marginal interfacial interactions. 
\end{abstract}



\maketitle

The equilibrium contact angle $\theta$ of a macroscopic drop of liquid on a planar substrate (wall) is determined by the tensions of the wall-gas, wall-liquid and liquid-gas interfaces, by Young's equation \cite{Rowlinson1982,Dietrich1988,Bonn2009}
\begin{equation}
\gamma_{wg}-\gamma_{wl}=\gamma_{lg}\cos\theta
\end{equation}
For complete wetting ($\theta=0$), the tensions satisfy Antonow's rule $\gamma_{wg}=\gamma_{wl}+\gamma_{lg}$, which means that, as the pressure is increased towards saturation, $p\to p_{sat}(T)$, at temperature $T$, a macroscopic layer of liquid must be adsorbed at the wall. However, for partial wetting ($\theta>0$), the wetting layer thickness remains finite at $p_{sat}$. It is well-known that fluid adsorption is strongly modified, and in general enhanced, by substrate geometry \cite{Quere2008,Finn1986,Rejmer1999,Parry2000a,Abraham2002,Milchev2003,Rascon2000a,Rascon2000,Dietrich2006,Bruschi2002,Ocko2005}. An example of this is the \textit{capillary condensation} of liquid in a slit of width $L$ at a shifted value of the pressure $p_{cc}(T;L)$ \cite{Evans1986,Evans1990a}, the details of which depend on whether the slit is capped at one end, thus forming a rectangular groove \cite{Umberto1989,Darbellay1992,Parry2007,Malijevsky2012,Rascon2013,
Yatsyshin2013}. Here, we point out that, in this 
groove geometry, one may identify a \textit{capillary} contact angle $\theta^\textup{cap}(L)$, defined by analogy with the Young equation but at capillary coexistence $p_{cc}$, rather than at bulk coexistence $p_{sat}$. This can be thought of as the angle at which the meniscus, separating capillary liquid and gas phases, meets the groove bottom as shown in Fig.~1. Intuitively, one may think that if all the walls are made of completely wet material ($\theta=0$), then the capillary contact angle is also zero ($\theta^\textup{cap}(L)=0$). This is indeed the case if all the walls are identical. However, if the bottom wall, which extends over the whole lower half-space, has a weaker long-ranged dispersion interaction with the fluid than the side walls, the capillary contact angle $\theta^\textup{cap}(L)$ is non-zero. Thus, while grooves made wholly of either material have $\theta^\textup{cap}(L)=0$, somewhat counter-intuitively, a groove made of a combination of both has $\theta^\textup{cap}(L)>0$. In addition, we show that $\theta^\textup{cap}(L)$ vanishes in two limits: a) as the slit becomes macroscopically wide, and ii) as $L$ is reduced to a specific value determined by the mismatch in Hamaker constants of the side and bottom walls.\\

\begin{figure}[t]
\includegraphics[width=.95\columnwidth]{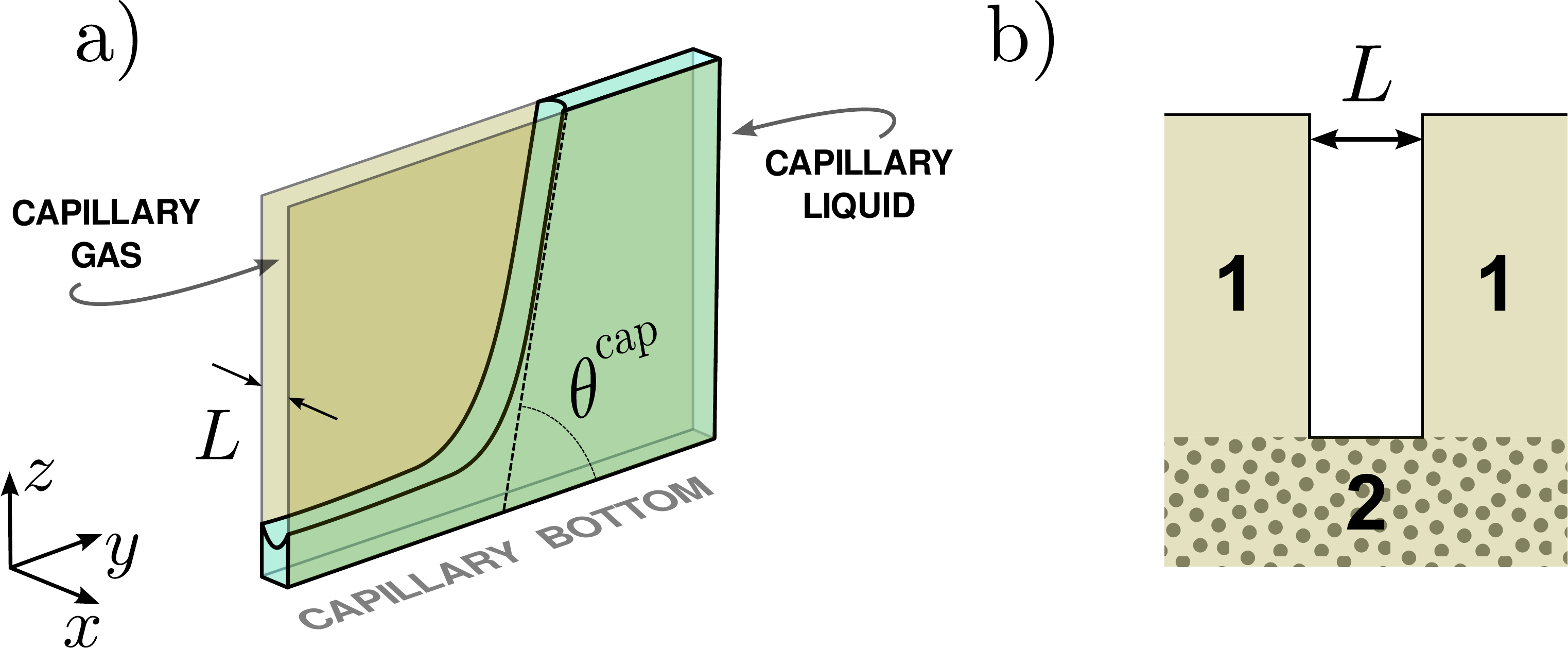}
\caption{a) Schematic illustration of a mesoscopic droplet of capillary liquid in a heterogeneous groove at capillary coexistence. The meniscus is of near circular cross-section, meeting the side walls tangentially, and forming an angle $\theta^\textup{cap}$ as it separates from the bottom. b) Cross-section of a heterogeneous groove made from two completely wet materials.}
\end{figure}

Consider the interface between a planar wall of infinite area, occupying the half-space $z<0$, and a bulk vapour at a subcritical temperature $T<T_c$ and pressure $p<p_{sat}(T)$ (or, equivalently, chemical potential $\mu<\mu_{sat}(T)$). If $\theta=0$, then as $p\to p_{sat}$, the equilibrium thickness $\ell_\pi$ of the adsorbed liquid layer grows and would become macroscopic in the absence of gravity. The divergence of $\ell_\pi$  was first understood by Frumkin and Derjaguin using the concept of a disjoining pressure \cite{Henderson2011}. Equivalently, one determines a binding potential $W_\pi(\ell)$ defined as the excess grand potential per unit area of a wetting film constrained to be of thickness $\ell$ \cite{Dietrich1988}. This quantity can be constructed from a microscopic density functional theory (DFT), where the Grand potential $\Omega[\rho]=F[\rho]-\int d{\bf{r}} \rho({\bf{r}})(\mu-V({\bf{r}}))$ is written as a functional of the one-body average density $\rho({\bf r})$. Here, $F[\rho]$ is the 
intrinsic Helmholtz functional modelling fluid-fluid interactions and $V({\bf{r}})$ is the external potential due to the 
wall(s) 
\cite{Evans1979}. Thus, for a single wall, $V({\bf{r}})=\rho_w\int  d{\bf{r}}'w(|{\bf{r}}-{\bf{r}}'|) $ where the integral is over the volume of the wall (of number density $\rho_w$) and  $w(r)$ is the pair potential between fluid and wall atoms. The binding potential then follows from $\Omega[\rho]$ using a sharp-kink approximation for the density profile $\rho({\bf{r}})$ in which one simply assumes that there is liquid of bulk density $\rho_l$ below the interface and bulk gas of density $\rho_g$ above it. For systems with dispersion forces whose wall-fluid and fluid-fluid  potentials decay proportional to $-\varepsilon^w r^{-6}$ and $-\varepsilon r^{-6}$, respectively, the binding potential has the well-known form \cite{Dietrich1988}
\begin{equation}
W_\pi(\ell)=\delta p\,\ell+\frac{A}{\ell^2}+\cdots
\end{equation}
where $\delta p=p_{sat}-p$. The first term is the thermodynamic penalty of having a layer of a metastable liquid. The second emerges after the interaction potentials are integrated over the 3D volume of the wall and the thickness of the wetting layer, and its coefficient identifies the Hamaker constant $A\propto (\rho_l-\rho_g)(\rho_w \varepsilon^w-\rho_l \varepsilon)$, which is positive for complete wetting. Minimization of $W_\pi(\ell)$ determines the equilibrium film thickness $\ell_\pi\sim \delta p ^{-1/3}$ \cite{Dietrich1988}. \\

Consider now a capillary groove of macroscopic length and depth but of microscopic width $L$ which is capped at its bottom. The groove is made from three slabs (two identical side walls and a bottom) of two different materials which are both completely wet. The side walls, of material 1 with interaction strength $\varepsilon_1^w$, occupy the regions $z>0$ and $|x|>L/2$. The slit is capped by having the third slab of material 2, with interaction strength $\varepsilon_2^w$, occupy the whole lower space, $z<0$ (See Fig.~1(b)). In practice, this can be achieved by depositing a layer of material 1 on material 2, and then etching a groove (or an array of them) whose width $L$ is much smaller than the material dimensions. The open end ($z\to \infty$) at the top of the capillary groove is in contact with a bulk gas at pressure $p$ and temperature $T$. In an uncapped slit, confinement between the side walls leads to the phenomenon of capillary-condensation corresponding to the shift of the bulk-like coexistence curve so that, at fixed $L$, capillary-liquid (CL) and capillary-gas (CG) phases coexist along a line $p_{cc}(T;L)$ which terminates at a capillary critical temperature $T_c(L)$. In the capped system, 
geometry necessitates the formation of a meniscus separating CL and CG phases at some distance $\ell_m$ from the bottom, which determines the adsorption $\Gamma\approx (\rho_l-\rho_g)L \ell_m$.

To find $\ell_m$ , we first consider a mean-field (MF) treatment and, using a sharp-kink approximation for the density profile, construct from $\Omega[\rho]$ a capillary binding potential $W_\text{cap} (L )$ by constraining the meniscus to a uniform height along the groove, and determine the excess grand potential per unit area of the groove bottom. If $\ell\gg L$, we find
\begin{equation}\label{Wcap}
W^\textup{cap}(\ell)\,=\,\Delta p\,\ell+\frac{A_2-A_1}{\ell^2}+\frac{3A_1 L}{8\ell^3}\cdots
\end{equation}
where $\Delta p= p_{cc}(T;L)-p$ and $A_1$, $A_2$ are the (positive) Hamaker constants for the side and bottom walls, respectively. The first term is the thermodynamic penalty of having a thick layer of CL and is analogous to the term $\delta p\,\ell$ in $W_\pi(\ell)$ except that pressure is now measured relative to capillary condensation. Analysis at this order also determines the value of $p_{sat}(T)-p_{cc}(T;L)=2\gamma_{lg}/(L-3\ell_\pi)$, which is the Kelvin-Derjaguin result for the shift of the coexistence line allowing for thick wetting films at the side walls \cite{Evans1990a}. The remaining terms in $W^\textup{cap}(\ell)$ arise from the dispersion forces and can be understood as follows: Consider an infinite uncapped capillary-slit exactly at $p=p_{cc}$, and place the meniscus at some arbitrary position. Now, cap the capillary by inserting an infinite slab of material type 1 of width $L$ at some large distance $\ell$ below the meniscus. 
Since the width of this slab is finite, the contribution to $\Omega[\rho]$ from the dispersion forces can only decay as $\mathcal{O}(\ell^{-3})$ (see the final term of Eq.~(\ref{Wcap})).
When we make the capillary heterogenous, we must further imagine slicing off an infinite slab of material 1 at the same depth and replacing it with an infinite slab of material type 2. The contribution to $\Omega[\rho]$ from the dispersion forces for both these slabs now involves integration over a 3D semi-volume, leading to the second term of Eq.~(\ref{Wcap}). We now consider three scenarios:\\

{\bf A}) \textit{A homogeneous capillary} ($A_1=A_2$). In this case, the meniscus is repelled from the capped end by a term of $\mathcal{O}(\ell^{-3})$, which competes with the thermodynamic attraction proportional to $\Delta p\,\ell$. Minimization of $W^\textup{cap}(\ell)$ determines the MF meniscus height $\ell_m\sim \Delta p^{-1/4}$ \cite{Parry2007}. The condensation occurring as $p\to p_{cc}^-$ is therefore a continuous capillary transition. \\

{\bf B}). \textit{A heterogeneous capillary} ($A_1<A_2$). Now, there is a stronger repulsion from the cap than in case {\bf A}, and the meniscus height grows as $\ell_m\sim \Delta p ^{-1/3}$, similar to complete wetting at a planar wall. The condensation transition remains continuous.\\

{\bf C}). \textit{A heterogeneous capillary} ($A_1>A_2$). Importantly, the difference between the Hamaker constants leads to an interfacial \textit{attraction}, so that the meniscus remains bound at a distance $\ell_m\approx 9L A_1/16 (A_1-A_2)$ from the cap, even at $p=p_{cc}$. This state coexists with one in which the groove is filled with CL.\\

The remarkable implication of this result is that, in a capillary with a less attractive bottom wall, the condensation transition is first-order even though it would be continuous in a homogeneous capillary made entirely of either material. We emphasise that this phenomenon only occurs if the less attractive bottom wall occupies the \textit{whole} lower half space (See Fig.~1(b)). If the slab of material type 2 capping the capillary occupies only the width of the slit, the effect is absent since the effective binding potential for this system is $W^\textup{cap}(\ell)=\Delta p\,\ell+3A_2 L/8\ell^3+\cdots$. The condensation remains continuous, as for case {\bf A}, albeit with a different amplitude.\\

\begin{figure}[t]
\includegraphics[width=.9\columnwidth]{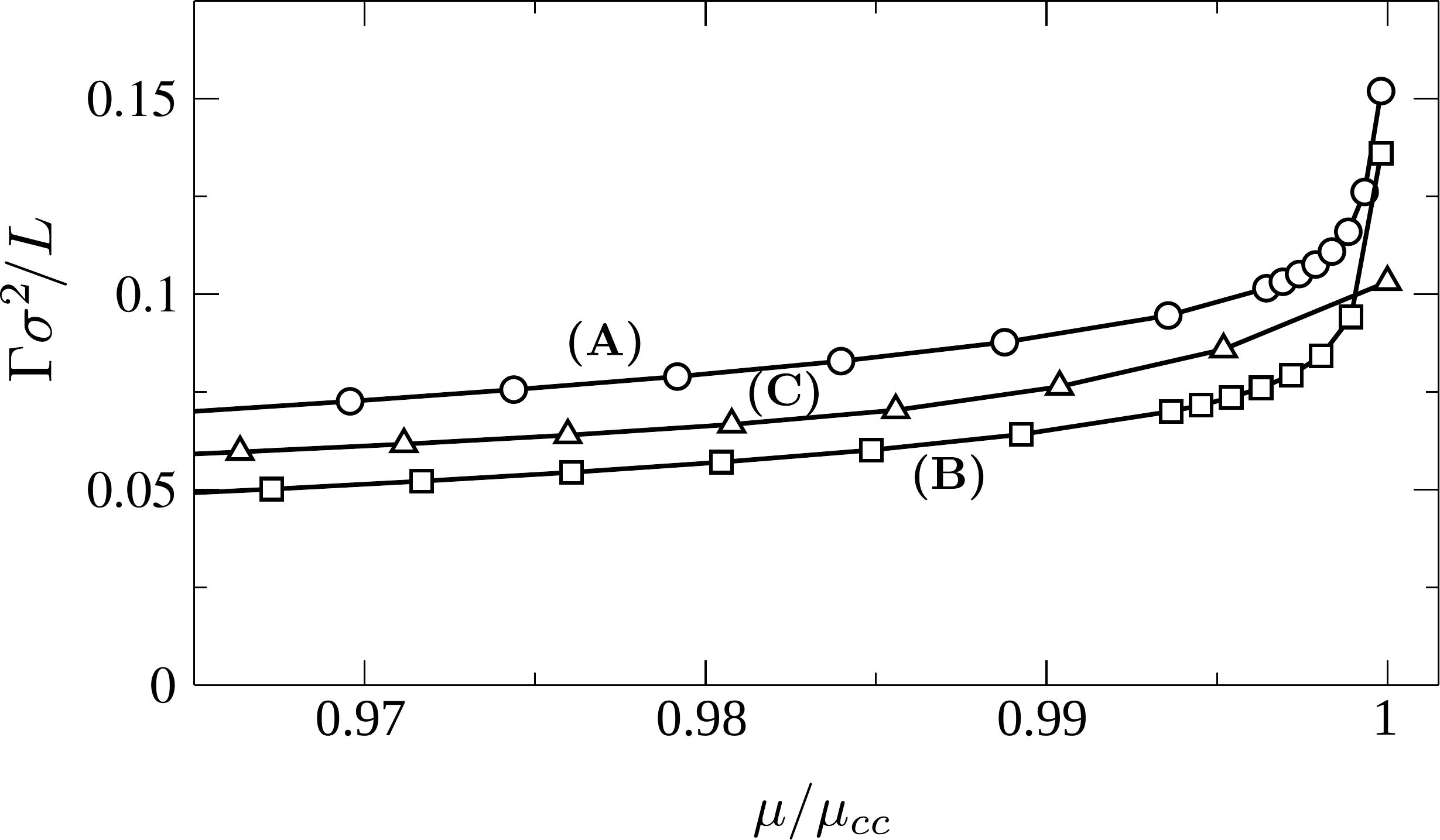}
\caption{Adsorption isotherms obtained by numerical minimization of the DFT for {\bf (A)} a homogeneous capillary, {\bf (B)} a capillary with a more attractive bottom wall, and {\bf (C)} a capillary with a less attractive bottom wall, for $L=12\sigma$ and $T=0.96T_c$. Here, $\mu_{cc}$ is the chemical potential at capillary condensation, determined independently for each infinite open slit.}
\end{figure}

We have tested these predictions using a Rosenfeld-like DFT \cite{Rosenfeld1989} with a mean-field treatment of the attractive fluid-fluid forces $F_{att}=\frac{1}{2}\int\!\!\int\! d{\bf r}_1 d{\bf r}_2\,\rho({\bf r}_1)\,u(r_{12})\,\rho({\bf r}_2)$. For the latter, we chose $u(r)=-4\varepsilon\,(\sigma/r)^{6}$, where $\sigma$ is the hard-sphere diameter. This attractive pair-potential is truncated at $r_c=2.5 \sigma$ and is set to zero inside the hard-sphere.

The external potential $V(x,z)$ has a hard-wall contribution and a long-ranged tail, which can be determined analytically from integrating the potential $-4\varepsilon^w_i\,(\sigma/r)^{6}$ over the volumes of the side ($i=1$) and bottom ($i=2$) walls. Far from the bottom of the capillary ($\sim 50\sigma$), we fix the density to that of a CG phase in order to model the open end of the groove. Translational invariance is assumed along the capillary (the $y$-axis). The temperature is set at $T=0.96\,T_c$ ($k_BT_c=1.41\varepsilon$), which is {\it above} the wetting temperatures of both the weaker ($\varepsilon^w=\varepsilon$, $T_w=0.93\,T_c$) and stronger ($\varepsilon^w=1.2\varepsilon$, $T_w=0.83\,T_c$) attractive walls, ensuring complete wetting of all surfaces. 

In Fig.~2, we show adsorption isotherms obtained from a full DFT calculation for three slits of width $L=12\sigma$: $\varepsilon_1^w=\varepsilon_2^w=1.2\varepsilon$ (case $\bf{A}$), $\varepsilon_1^w=\varepsilon$ and $\varepsilon_2^w=1.2\varepsilon$ (case $\bf{B}$), and $\varepsilon_1^w=1.2 \varepsilon$ and $\varepsilon_2^w=\varepsilon$ (case $\bf{C}$). As predicted, the condensation is continuous for the first two cases, and a log-log plot shows very good agreement with the predicted exponent values $1/4$ and $1/3$, respectively (See Fig.~3a). For the third case, with a less attractive bottom wall, the condensation transition is first-order and, at capillary coexistence, a meniscus remains bound close to the cap (Fig.~3b).\\

Just as Young's equation allows us to define a contact angle $\theta$ from the three surface tensions associated with coexisting bulk phases and a single isolated wall at $p=p_{sat}$, we may now define a \textit{capillary} contact angle from the analogous free energies of the coexisting capillary phases at $p=p_{cc}$:

\begin{equation}
\label{capyoung}
\gamma_{wg}^\textup{cap}(L)-\gamma_{wl}^\textup{cap}(L)\;=\;\gamma_{lg}^\textup{cap}(L)\,\cos\theta^\textup{cap}(L)
\end{equation}

\begin{figure}[t]
\includegraphics[width=\columnwidth]{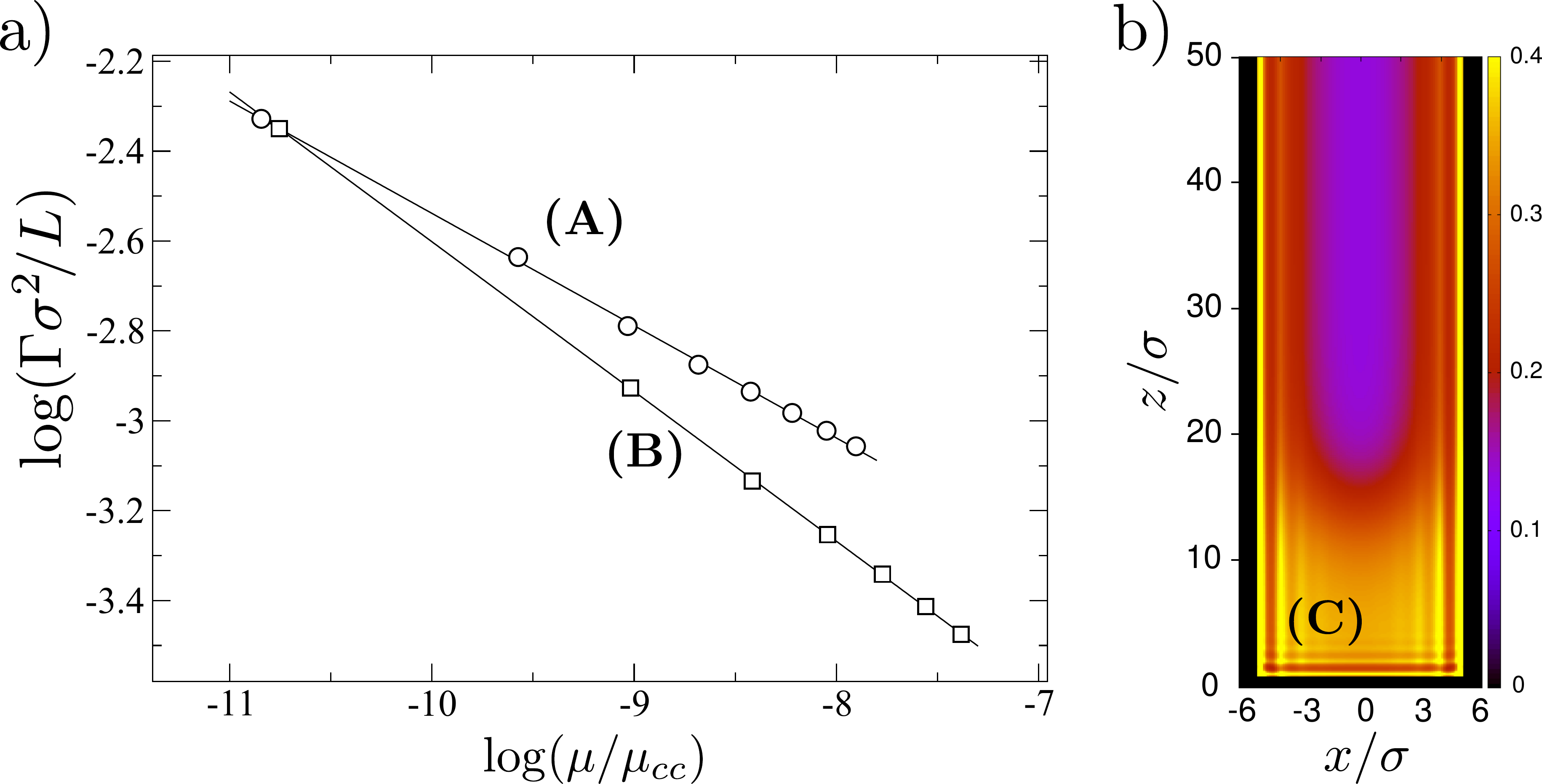}
\caption{a) Log-log plot of adsorption isotherms for the two examples of continuous capillary condensation, and comparison with the predicted slopes $-1/4$ and $-1/3$ (cases {\bf{(A)}} and {\bf{(B)}}, respectively). b) For case {\bf(C)}, a two dimensional density profile $\rho(x,z)$ showing a bound state meniscus configuration that coexists with a completely filled capillary at $\mu=\mu_{cc}$.}
\end{figure}

Here, $\gamma_{lg}^\textup{cap}(L)$ is the surface tension associated with the meniscus separating capillary liquid and capillary gas phases, defined as the excess grand potential per unit area of the capillary bottom. For wide slits, this tension is well approximated by $\gamma_{lg}^\textup{cap}(L)\approx \pi \gamma_{lg}/2$, owing to the near circular shape of the meniscus. Similarly, $\gamma_{wg}^\textup{cap}(L)$ and $\gamma_{wl}^\textup{cap}(L)$ are the surface tensions associated with the interface between the groove bottom and the CG phase (bound meniscus) and CL phase (unbound meniscus), respectively.

At MF level, we can identify $W^\textup{cap}(\ell_m)=\gamma_{lg}^\textup{cap}(L)\left(\cos\theta^\textup{cap}(L)-1\right)$, which leads to 
\begin{equation}
\theta^\textup{cap}(L)\;\approx\;\frac{\;32\,(A_1\!-\!A_2)^{\frac{3}{2}}}{9\sqrt{3\pi\gamma_{lg}}\,A_1 L};\qquad L\to\infty,
\label{thetacMF}
\end{equation}
valid for $A_1>A_2$ and sufficiently large $L$. Otherwise, when $A_1\le A_2$, the capillary contact angle vanishes.\\ 

The MF result (\ref{thetacMF}) suggests that we can induce a capillary wetting transition by changing the sign of $A_1-A_2$, similar to the standard mechanism for the critical wetting transition at a single planar wall \cite{Dietrich1988,Bonn2009}. However, rather than tuning the Hamaker constants, we focus instead on how $\theta^\textup{cap}(L)$ depends on $L$, while maintaining capillary coexistence $p=p_{cc}(L)$. To do this, we must go beyond MF and consider fluctuation effects arising from the wandering of the meniscus height along the groove ($y$ axis). These are well described by the 1D interfacial Hamiltonian 
\begin{equation}
H^\textup{cap}[\ell]\;=\;L\int\!\!dy\;\left\{\frac{\gamma_{lg}^\textup{cap}(L)}{2} \left(\frac{d\ell}{dy}\right)^2+W^\textup{cap}(\ell)\right\}
\end{equation}
where $\ell(y)$ denotes the local height of the meniscus at position $y$, and one may approximate $\gamma_{lg}^\textup{cap}(L)\approx \pi \gamma_{lg}/2$.
The partition function can be evaluated exactly using standard transfer-matrix techniques, the spectrum of which follows from solution of a Schr\"odinger-like equation from which one can readily determine $\ell_m=\langle\,\ell\,\rangle$, the roughness $\xi_\perp=\sqrt{\langle\,\ell^2\,\rangle-\ell_m^2}$ and the lengthscale $\xi_y$ describing height correlations along the direction of the groove. Analysis shows that $\theta^\textup{cap}$ may vanish in two different ways. The first occurs when the slit becomes macroscopically wide, in which case interfacial/meniscus fluctuations are suppressed. Thus, $\theta^\textup{cap}(L)$ vanishes according to (\ref{thetacMF}) with the accompanying scaling behaviour $\ell_m\sim L$, $\xi_\perp\sim \sqrt{L}$ and $\xi_y\sim L^2$ describing the growth of the meniscus.\\

The second type of capillary transition involving the meniscus occurs as the slit width decreases. According to the MF result (\ref{thetacMF}), the capillary contact angle $\theta^\textup{cap}(L)$ continues to increase as the width becomes microscopic. However, the reduction in the stiffness coefficient $L\,\gamma_{lg}^\textup{cap}(L)$ enhances fluctuation effects, and the meniscus eventually tunnels out of the potential well in $W^\textup{cap}$. Thus, at a sufficiently small slit separation $L=L_w$, the capillary contact angle $\theta^\textup{cap}$ also vanishes, corresponding to another capillary wetting transition. This transition belongs to the intermediate fluctuation regime of two dimensional critical wetting, because the $\ell^{-2}$ interaction is marginal, making it highly sensitive to the short-ranged structure of the binding potential \cite{Lipowsky1988}. In our case, the final term in $W^\textup{cap}(\ell)$ (see Eq.~(\ref{Wcap})) is strongly repulsive, which means that the transition is 
characterised by essential singularities \cite{Chui1983,Kroll1983}. 
When the difference in the Hamaker constants is small, this identifies the value of the slit width $L_w$ at which $\theta^\textup{cap}(L)$ vanishes as
\begin{equation}
L_w=\frac{k_B T}{2\sqrt{\pi\gamma_{lg}(A_1-A_2)}},\qquad A_1>A_2.
\end{equation}
Note that the divergence of $L_w$ as $A_2\to A_1$ is consistent with the fact that this transition is absent in a homogeneous capillary. When $A_1=A_2$, the capillary contact angle is always zero. As $L$ is decreased towards $L_{w}$ in a heterogeneous capillary, the capillary contact angle vanishes as
\begin{equation}\label{thetacFF}
\theta^\textup{cap}(L)\;\sim\; e^{-\frac{2\pi L_{w}}{\sqrt{L^2-L_{w}^2}}};\qquad L\to L_w
\end{equation}
with the accompanying scaling $\ell_m\sim\xi_\perp\sim\xi_y^{1/2}\propto 1/\theta^\textup{cap}$, characteristic of fluctuation-dominated behaviour.
For narrower grooves ($L<L_w$), complete wetting of the cap is restored ($\theta^\textup{cap}=0$) and eventually coexistence ends at a conventional capillary critical point \cite{Evans1986}. 
These features are illustrated schematically in Fig.~4, where we plot $\theta^\textup{cap}$ vs.~$\tau\equiv\sqrt{k_B T/4\pi\gamma_{lg} L^2\,}$. This dimensionless parameter may be interpreted in two ways. At fixed $T<T_c$, increasing $\tau$ corresponds to decreasing $L$ to the critical slit width $L_c(T)$, at which capillary coexistence between CL and CG phases ends. Alternatively, at fixed $L$, increasing $\tau$ corresponds to increasing $T$ towards the capillary critical temperature $T_c(L)$. 
The value of $\tau$ at the capillary critical point depends on the slit width but, in the limit $L\to\infty$, tends to a universal value $\tau_c$. Using the known values of the critical amplitude ratios associated with the wetting parameter \cite{Evans1992} and critical point shift $T_c(L)-T_c$ \cite{Nakanishi1983}, this can be reliably estimated as $\tau_c\approx 0.1$. 
The vanishing of $\theta^\textup{cap}$, as described by (\ref{thetacMF}) and (\ref{thetacFF}), corresponds to the two different capillary wetting transitions, which occur at $\tau=0$ and $\tau=\tau_w\equiv\sqrt{\frac{A_1-A_2}{k_BT}}$, respectively. The maximum value of $\theta^\textup{cap}$ occurs between these two transitions and is of order $A_1/k_B T$ if the difference between the Hamaker constants is large.\\

In summary, we have shown that, in a capillary groove, the competition between the wall-fluid dispersion forces at the bottom and side walls can lead to a non-zero capillary contact angle, though the isolated walls exhibit complete wetting. This finite $\theta^\textup{cap}$ will be present for all temperatures away from the near vicinity of the capillary critical point if the mismatch between the Hamaker constants is of order $k_BT$. Thus, even though the wetting transitions at $\tau=0$ and $\tau=\tau_w$ may be difficult to observe experimentally, the qualitative change to the order of capillary condensation should be readily observable in grooves of micron size, very similar to the experiments of Mistura et al.\ reported in \cite{Rascon2013}.\\

\begin{figure}[h]
\includegraphics[width=0.9\columnwidth]{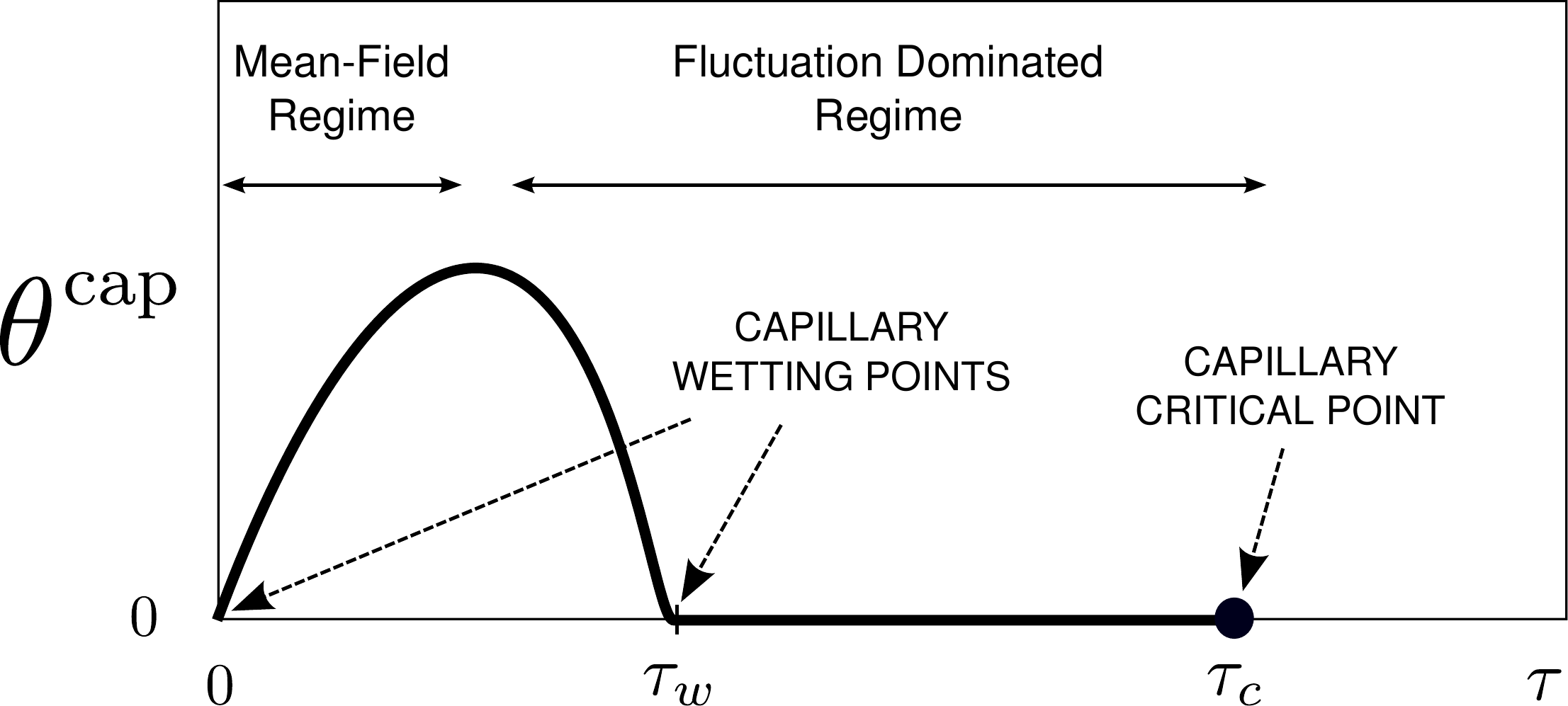}
\caption{Schematic behaviour of the capillary contact angle $\theta^\textup{cap}$ as a function of the dimensionless variable $\tau\equiv\sqrt{k_B T/4\pi\gamma_{lg} L^2\,}$. The locations of the capillary wetting transitions and capillary critical point are shown.}
\end{figure}

\acknowledgments

A.O.P.~wishes to thank the support of the EPSRC UK for grant EP/J009636/1. A.M.~thanks the Czech Science Foundation for grant 13-09914S. C.R.~acknowledges support from grants FIS2010-22047-C05 and MODELICO.\\

\bibliography{wetting}

\begin{thebibliography}{30}
\expandafter\ifx\csname natexlab\endcsname\relax\def\natexlab#1{#1}\fi
\expandafter\ifx\csname bibnamefont\endcsname\relax
  \def\bibnamefont#1{#1}\fi
\expandafter\ifx\csname bibfnamefont\endcsname\relax
  \def\bibfnamefont#1{#1}\fi
\expandafter\ifx\csname citenamefont\endcsname\relax
  \def\citenamefont#1{#1}\fi
\expandafter\ifx\csname url\endcsname\relax
  \def\url#1{\texttt{#1}}\fi
\expandafter\ifx\csname urlprefix\endcsname\relax\def\urlprefix{URL }\fi
\providecommand{\bibinfo}[2]{#2}
\providecommand{\eprint}[2][]{\url{#2}}

\bibitem[{\citenamefont{Rowlinson and Widom}(1982)}]{Rowlinson1982}
\bibinfo{author}{\bibfnamefont{J.~S.} \bibnamefont{Rowlinson}}
  \bibnamefont{and} \bibinfo{author}{\bibfnamefont{B.}~\bibnamefont{Widom}},
  \emph{\bibinfo{title}{Molecular Theory of Capillarity}}
  (\bibinfo{publisher}{Clarendon Press}, \bibinfo{year}{1982}).

\bibitem[{\citenamefont{Dietrich}(1988)}]{Dietrich1988}
\bibinfo{author}{\bibfnamefont{S.}~\bibnamefont{Dietrich}}, in
  \emph{\bibinfo{booktitle}{Phase Transitions and Critical Phenomena}}, edited
  by \bibinfo{editor}{\bibfnamefont{C.}~\bibnamefont{Domb}} \bibnamefont{and}
  \bibinfo{editor}{\bibfnamefont{J.}~\bibnamefont{Lebowitz}}
  (\bibinfo{publisher}{Academic Press Limited}, \bibinfo{year}{1988}),
  vol.~\bibinfo{volume}{12}.

\bibitem[{\citenamefont{Bonn et~al.}(2009)\citenamefont{Bonn, Eggers, Indekeu,
  Meunier, and Rolley}}]{Bonn2009}
\bibinfo{author}{\bibfnamefont{D.}~\bibnamefont{Bonn}},
  \bibinfo{author}{\bibfnamefont{J.}~\bibnamefont{Eggers}},
  \bibinfo{author}{\bibfnamefont{J.~O.} \bibnamefont{Indekeu}},
  \bibinfo{author}{\bibfnamefont{J.}~\bibnamefont{Meunier}}, \bibnamefont{and}
  \bibinfo{author}{\bibfnamefont{E.}~\bibnamefont{Rolley}},
  \bibinfo{journal}{Rev. Mod. Phys.} \textbf{\bibinfo{volume}{81}},
  \bibinfo{pages}{739} (\bibinfo{year}{2009}).

\bibitem[{\citenamefont{Qu{\'e}r{\'e}}(2008)}]{Quere2008}
\bibinfo{author}{\bibfnamefont{D.}~\bibnamefont{Qu{\'e}r{\'e}}},
  \bibinfo{journal}{Annu. Rev. Mater. Res.} \textbf{\bibinfo{volume}{38}},
  \bibinfo{pages}{71} (\bibinfo{year}{2008}).

\bibitem[{\citenamefont{Finn}(1986)}]{Finn1986}
\bibinfo{author}{\bibfnamefont{R.}~\bibnamefont{Finn}},
  \emph{\bibinfo{title}{Equilibrium Capillary Surfaces}}
  (\bibinfo{publisher}{Springer}, \bibinfo{year}{1986}).

\bibitem[{\citenamefont{Rejmer et~al.}(1999)\citenamefont{Rejmer, Dietrich, and
  Napi{\'o}rkowski}}]{Rejmer1999}
\bibinfo{author}{\bibfnamefont{K.}~\bibnamefont{Rejmer}},
  \bibinfo{author}{\bibfnamefont{S.}~\bibnamefont{Dietrich}}, \bibnamefont{and}
  \bibinfo{author}{\bibfnamefont{M.}~\bibnamefont{Napi{\'o}rkowski}},
  \bibinfo{journal}{Phys. Rev. E} \textbf{\bibinfo{volume}{60}},
  \bibinfo{pages}{4027} (\bibinfo{year}{1999}).

\bibitem[{\citenamefont{Parry et~al.}(2000)\citenamefont{Parry, Rasc{\'o}n, and
  Wood}}]{Parry2000a}
\bibinfo{author}{\bibfnamefont{A.~O.} \bibnamefont{Parry}},
  \bibinfo{author}{\bibfnamefont{C.}~\bibnamefont{Rasc{\'o}n}},
  \bibnamefont{and} \bibinfo{author}{\bibfnamefont{A.~J.} \bibnamefont{Wood}},
  \bibinfo{journal}{Phys. Rev. Lett.} \textbf{\bibinfo{volume}{85}},
  \bibinfo{pages}{345} (\bibinfo{year}{2000}).

\bibitem[{\citenamefont{Abraham and Maciolek}(2002)}]{Abraham2002}
\bibinfo{author}{\bibfnamefont{D.~B.} \bibnamefont{Abraham}} \bibnamefont{and}
  \bibinfo{author}{\bibfnamefont{A.}~\bibnamefont{Maciolek}},
  \bibinfo{journal}{Phys. Rev. Lett.} \textbf{\bibinfo{volume}{89}},
  \bibinfo{pages}{286101} (\bibinfo{year}{2002}).

\bibitem[{\citenamefont{Milchev et~al.}(2003)\citenamefont{Milchev, M{\"u}ller,
  Binder, and Landau}}]{Milchev2003}
\bibinfo{author}{\bibfnamefont{A.}~\bibnamefont{Milchev}},
  \bibinfo{author}{\bibfnamefont{M.}~\bibnamefont{M{\"u}ller}},
  \bibinfo{author}{\bibfnamefont{K.}~\bibnamefont{Binder}}, \bibnamefont{and}
  \bibinfo{author}{\bibfnamefont{D.~P.} \bibnamefont{Landau}},
  \bibinfo{journal}{Phys. Rev. Lett.} \textbf{\bibinfo{volume}{90}},
  \bibinfo{pages}{136101} (\bibinfo{year}{2003}).

\bibitem[{\citenamefont{Rasc{\'o}n and
  Parry}(2000{\natexlab{a}})}]{Rascon2000a}
\bibinfo{author}{\bibfnamefont{C.}~\bibnamefont{Rasc{\'o}n}} \bibnamefont{and}
  \bibinfo{author}{\bibfnamefont{A.~O.} \bibnamefont{Parry}},
  \bibinfo{journal}{J. Chem. Phys.} \textbf{\bibinfo{volume}{112}},
  \bibinfo{pages}{5175} (\bibinfo{year}{2000}{\natexlab{a}}).

\bibitem[{\citenamefont{Rasc{\'o}n and Parry}(2000{\natexlab{b}})}]{Rascon2000}
\bibinfo{author}{\bibfnamefont{C.}~\bibnamefont{Rasc{\'o}n}} \bibnamefont{and}
  \bibinfo{author}{\bibfnamefont{A.~O.} \bibnamefont{Parry}},
  \bibinfo{journal}{Nature} \textbf{\bibinfo{volume}{407}},
  \bibinfo{pages}{986} (\bibinfo{year}{2000}{\natexlab{b}}).

\bibitem[{\citenamefont{Dietrich and Tasinkevych}(2006)}]{Dietrich2006}
\bibinfo{author}{\bibfnamefont{S.}~\bibnamefont{Dietrich}} \bibnamefont{and}
  \bibinfo{author}{\bibfnamefont{M.}~\bibnamefont{Tasinkevych}},
  \bibinfo{journal}{Phys. Rev. Lett.} \textbf{\bibinfo{volume}{97}},
  \bibinfo{pages}{106102} (\bibinfo{year}{2006}).

\bibitem[{\citenamefont{Bruschi et~al.}(2002)\citenamefont{Bruschi, Carlin, and
  Mistura}}]{Bruschi2002}
\bibinfo{author}{\bibfnamefont{L.}~\bibnamefont{Bruschi}},
  \bibinfo{author}{\bibfnamefont{A.}~\bibnamefont{Carlin}}, \bibnamefont{and}
  \bibinfo{author}{\bibfnamefont{G.}~\bibnamefont{Mistura}},
  \bibinfo{journal}{Phys. Rev. Lett.} \textbf{\bibinfo{volume}{89}},
  \bibinfo{pages}{166101} (\bibinfo{year}{2002}).

\bibitem[{\citenamefont{Gang et~al.}(2005)\citenamefont{Gang, Alvine, Fukuto,
  Pershan, Black, and Ocko}}]{Ocko2005}
\bibinfo{author}{\bibfnamefont{O.}~\bibnamefont{Gang}},
  \bibinfo{author}{\bibfnamefont{K.~J.} \bibnamefont{Alvine}},
  \bibinfo{author}{\bibfnamefont{M.}~\bibnamefont{Fukuto}},
  \bibinfo{author}{\bibfnamefont{P.~S.} \bibnamefont{Pershan}},
  \bibinfo{author}{\bibfnamefont{C.~T.} \bibnamefont{Black}}, \bibnamefont{and}
  \bibinfo{author}{\bibfnamefont{B.~M.} \bibnamefont{Ocko}},
  \bibinfo{journal}{Phys. Rev. Lett.} \textbf{\bibinfo{volume}{95}},
  \bibinfo{pages}{217801} (\bibinfo{year}{2005}).

\bibitem[{\citenamefont{Evans et~al.}(1986)\citenamefont{Evans, {Marini Bettolo
  Marconi}, and Tarazona}}]{Evans1986}
\bibinfo{author}{\bibfnamefont{R.}~\bibnamefont{Evans}},
  \bibinfo{author}{\bibfnamefont{U.}~\bibnamefont{{Marini Bettolo Marconi}}},
  \bibnamefont{and} \bibinfo{author}{\bibfnamefont{P.}~\bibnamefont{Tarazona}},
  \bibinfo{journal}{J. Chem. Phys.} \textbf{\bibinfo{volume}{84}},
  \bibinfo{pages}{2376} (\bibinfo{year}{1986}).

\bibitem[{\citenamefont{Evans}(1990)}]{Evans1990a}
\bibinfo{author}{\bibfnamefont{R.}~\bibnamefont{Evans}}, \bibinfo{journal}{J.
  Phys.: Condens. Matter} \textbf{\bibinfo{volume}{2}}, \bibinfo{pages}{8989}
  (\bibinfo{year}{1990}).

\bibitem[{\citenamefont{Marconi and Van~Swol}(1989)}]{Umberto1989}
\bibinfo{author}{\bibfnamefont{U.~M.~B.} \bibnamefont{Marconi}}
  \bibnamefont{and} \bibinfo{author}{\bibfnamefont{F.}~\bibnamefont{Van~Swol}},
  \bibinfo{journal}{Phys. Rev. E} \textbf{\bibinfo{volume}{39}},
  \bibinfo{pages}{4109} (\bibinfo{year}{1989}).

\bibitem[{\citenamefont{Darbellay and Yeomans}(1992)}]{Darbellay1992}
\bibinfo{author}{\bibfnamefont{G.~A.} \bibnamefont{Darbellay}}
  \bibnamefont{and} \bibinfo{author}{\bibfnamefont{J.~M.}
  \bibnamefont{Yeomans}}, \bibinfo{journal}{J. Phys.: Condens. Matter}
  \textbf{\bibinfo{volume}{25}}, \bibinfo{pages}{4275} (\bibinfo{year}{1992}).

\bibitem[{\citenamefont{Parry et~al.}(2007)\citenamefont{Parry, Rasc{\'o}n,
  Wilding, and Evans}}]{Parry2007}
\bibinfo{author}{\bibfnamefont{A.~O.} \bibnamefont{Parry}},
  \bibinfo{author}{\bibfnamefont{C.}~\bibnamefont{Rasc{\'o}n}},
  \bibinfo{author}{\bibfnamefont{N.~B.} \bibnamefont{Wilding}},
  \bibnamefont{and} \bibinfo{author}{\bibfnamefont{R.}~\bibnamefont{Evans}},
  \bibinfo{journal}{Phys. Rev. Lett.} \textbf{\bibinfo{volume}{98}},
  \bibinfo{pages}{226101} (\bibinfo{year}{2007}).

\bibitem[{\citenamefont{Malijevsk{\'y}}(2012)}]{Malijevsky2012}
\bibinfo{author}{\bibfnamefont{A.}~\bibnamefont{Malijevsk{\'y}}},
  \bibinfo{journal}{J. Chem. Phys} \textbf{\bibinfo{volume}{137}},
  \bibinfo{pages}{214704} (\bibinfo{year}{2012}).

\bibitem[{\citenamefont{Rasc{\'o}n et~al.}(2013)\citenamefont{Rasc{\'o}n,
  Parry, N{\"u}renberg, Pozatto, Tormen, Bruschi, and Mistura}}]{Rascon2013}
\bibinfo{author}{\bibfnamefont{C.}~\bibnamefont{Rasc{\'o}n}},
  \bibinfo{author}{\bibfnamefont{A.~O.} \bibnamefont{Parry}},
  \bibinfo{author}{\bibfnamefont{R.}~\bibnamefont{N{\"u}renberg}},
  \bibinfo{author}{\bibfnamefont{A.}~\bibnamefont{Pozatto}},
  \bibinfo{author}{\bibfnamefont{M.}~\bibnamefont{Tormen}},
  \bibinfo{author}{\bibfnamefont{L.}~\bibnamefont{Bruschi}}, \bibnamefont{and}
  \bibinfo{author}{\bibfnamefont{G.}~\bibnamefont{Mistura}},
  \bibinfo{journal}{J. Phys.: Condens. Matter} \textbf{\bibinfo{volume}{25}},
  \bibinfo{pages}{192101} (\bibinfo{year}{2013}).

\bibitem[{\citenamefont{Yatsyshin et~al.}(2013)\citenamefont{Yatsyshin, Savva,
  and Kalliadasis}}]{Yatsyshin2013}
\bibinfo{author}{\bibfnamefont{P.}~\bibnamefont{Yatsyshin}},
  \bibinfo{author}{\bibfnamefont{N.}~\bibnamefont{Savva}}, \bibnamefont{and}
  \bibinfo{author}{\bibfnamefont{S.}~\bibnamefont{Kalliadasis}},
  \bibinfo{journal}{Phys. Rev. E} \textbf{\bibinfo{volume}{87}},
  \bibinfo{pages}{020402(R)} (\bibinfo{year}{2013}).

\bibitem[{\citenamefont{Henderson}(2011)}]{Henderson2011}
\bibinfo{author}{\bibfnamefont{J.~R.} \bibnamefont{Henderson}},
  \bibinfo{journal}{Eur. Phys. J. Special Topics}
  \textbf{\bibinfo{volume}{197}}, \bibinfo{pages}{115} (\bibinfo{year}{2011}).

\bibitem[{\citenamefont{Evans}(1979)}]{Evans1979}
\bibinfo{author}{\bibfnamefont{R.}~\bibnamefont{Evans}}, \bibinfo{journal}{Adv.
  Phys.} \textbf{\bibinfo{volume}{28}}, \bibinfo{pages}{143}
  (\bibinfo{year}{1979}).

\bibitem[{\citenamefont{Rosenfeld}(1989)}]{Rosenfeld1989}
\bibinfo{author}{\bibfnamefont{Y.}~\bibnamefont{Rosenfeld}},
  \bibinfo{journal}{Phys. Rev. Lett.} \textbf{\bibinfo{volume}{63}},
  \bibinfo{pages}{980} (\bibinfo{year}{1989}).

\bibitem[{\citenamefont{Lipowsky and Nieuwenhuizen}(1988)}]{Lipowsky1988}
\bibinfo{author}{\bibfnamefont{R.}~\bibnamefont{Lipowsky}} \bibnamefont{and}
  \bibinfo{author}{\bibfnamefont{T.~M.} \bibnamefont{Nieuwenhuizen}},
  \bibinfo{journal}{J. Phys. A: Math. Gen.} \textbf{\bibinfo{volume}{21}},
  \bibinfo{pages}{L89} (\bibinfo{year}{1988}).

\bibitem[{\citenamefont{Chui and Ma}(1983)}]{Chui1983}
\bibinfo{author}{\bibfnamefont{S.~T.} \bibnamefont{Chui}} \bibnamefont{and}
  \bibinfo{author}{\bibfnamefont{K.~B.} \bibnamefont{Ma}},
  \bibinfo{journal}{Phys. Rev. B} \textbf{\bibinfo{volume}{28}},
  \bibinfo{pages}{2555} (\bibinfo{year}{1983}).

\bibitem[{\citenamefont{Kroll and Lipowsky}(1983)}]{Kroll1983}
\bibinfo{author}{\bibfnamefont{D.~M.} \bibnamefont{Kroll}} \bibnamefont{and}
  \bibinfo{author}{\bibfnamefont{R.}~\bibnamefont{Lipowsky}},
  \bibinfo{journal}{Phys. Rev. B} \textbf{\bibinfo{volume}{28}},
  \bibinfo{pages}{5273} (\bibinfo{year}{1983}).

\bibitem[{\citenamefont{Evans et~al.}(1992)\citenamefont{Evans, Hoyle, and
  Parry}}]{Evans1992}
\bibinfo{author}{\bibfnamefont{R.}~\bibnamefont{Evans}},
  \bibinfo{author}{\bibfnamefont{D.~C.} \bibnamefont{Hoyle}}, \bibnamefont{and}
  \bibinfo{author}{\bibfnamefont{A.~O.} \bibnamefont{Parry}},
  \bibinfo{journal}{Phys. Rev. A} \textbf{\bibinfo{volume}{45}},
  \bibinfo{pages}{3823} (\bibinfo{year}{1992}).

\bibitem[{\citenamefont{Nakanishi and Fisher}(1983)}]{Nakanishi1983}
\bibinfo{author}{\bibfnamefont{H.}~\bibnamefont{Nakanishi}} \bibnamefont{and}
  \bibinfo{author}{\bibfnamefont{M.~E.} \bibnamefont{Fisher}},
  \bibinfo{journal}{J. Chem. Phys.} \textbf{\bibinfo{volume}{78}},
  \bibinfo{pages}{3279} (\bibinfo{year}{1983}).

\end{thebibliography}

\end{document}